\documentclass[pra,aps,twocolumn,10pt,showpacs,nofootinbib]{revtex4-1}
\usepackage[utf8]{inputenc}
\usepackage[T1]{fontenc}
\usepackage{amsmath}
\usepackage{amsfonts}
\usepackage{amssymb}
\usepackage{graphicx} 
\usepackage{bm}
\newcommand{\mi}{\mathrm{i}}
\newcommand{\me}{\mathrm{e}}

\begin{document}

\title{Solving the quantum master equation of coupled harmonic oscillators with 
Lie algebra methods}
\author{Lucas Teuber}
\author{Stefan Scheel}
\email{stefan.scheel@uni-rostock.de}
\affiliation{Institut f\"ur Physik, Universit\"at Rostock, 
Albert-Einstein--Stra{\ss}e 23-24, D-18059 Rostock, Germany}

\date{\today}

\begin{abstract}
Based on a Liouville-space formulation of open systems, we present two methods 
to solve the quantum dynamics of coupled harmonic oscillators experiencing 
Markovian loss. Starting point is the quantum master equation in Liouville 
space which is generated by a Liouvillian that induces a Lie algebra. We show 
how this Lie algebra allows to define ladder operators that construct Fock-like 
eigenstates of the Liouvillian. These eigenstates are used to decompose the 
time-evolved density matrix and, together with the accompanying eigenvalues, 
provide insight into the transport properties of the lossy system. 
Additionally, a Wei-Norman expansion of the generated time evolution can be 
found by a structure analysis of the algebra. This structure analysis yields a 
construction principle to implement effective non-Hermitian Hamiltonians in 
lossy systems.
\end{abstract}

% \pacs{????}

\maketitle

%%%%%%%%%%%%%%%%%%%%%%%%%%%%%%%%%%%%%%%%%%%%%%%%%%%%%%%%%%%%%%%%%%%%%%
\section{Introduction}

The theory of open quantum systems developed from a bare necessity of 
describing realistic quantum systems to a tool for their deliberate design.
As a result, the non-Hermiticity of open systems was no longer deemed a hindrance 
but became a potential resource of interesting effects in its own right.
In recent years, interest in such 
non-Hermitian systems soared spanning from investigations of parity-time 
($\mathcal{PT}$) symmetry \cite{BenderBoettcher}, which preserves the reality of 
the spectrum of the Hamiltonian, over non-Hermitian coalescence of modes at 
exceptional points (EP) \cite{Heiss}, to fundamental questions regarding 
generalisations to non-Hermitian quantum mechanics \cite{Brody}.

Experimental realisations of $\mathcal{PT}$ symmetry or exceptional points can 
be found especially in optical systems \cite{Miri2019,Rueter,Peng} due to their high 
level of control over the gain and loss, but other implementations exist such as
LC circuits \cite{Schindler}, microwave cavities \cite{Dembowski}, or atoms \cite{Zhang}. 
Usually, these implementations are based on classical wave mechanics that 
account for first-quantisation effects where loss and gain are modelled by an 
effective non-Hermitian Hamiltonian. For the early research this was sufficient 
and resulted in many interesting discoveries such as changes of transport 
behaviour in broken and unbroken $\mathcal{PT}$ symmetry \cite{Makris} or 
increased measurement sensitivity at EPs \cite{Hodaei}.

However, recently efforts have been made to also investigate 
second-quantisation effects in non-Hermitian systems with one successful 
example being the measurement of two-photon Hong-Ou-Mandel correlations in a 
$\mathcal{PT}$-symmetric waveguide coupler \cite{Klauck19}. The key step was 
the use of passive $\mathcal{PT}$ symmetry meaning that the required 
antisymmetric gain/loss distribution was replaced by an all-loss distribution 
with the same antisymmetry. This passive scheme was used because any gain added 
to the system also adds noise that irrevocably changes the quantum state under 
study which negatively affects the observance of any $\mathcal{PT}$ quantum 
effect \cite{Scheel}. Thus, the replacement of gain/loss structures by pure 
loss structures seems to be one way to implement non-Hermitian quantum dynamics.

In this work, we present a theoretical description of the quantum dynamics 
of lossy quantum systems that lays the foundations to further understand the 
quantum effects occurring in non-Hermitian systems which might pave the way to 
investigate and apply quantum $\mathcal{PT}$ symmetry and quantum EPs.
For this, we present two methods to obtain the dynamics of coupled quantum
harmonic oscillators experiencing Markovian losses. Such oscillator systems 
model lossy waveguide arrays but are also applicable to other systems of 
bosonic nature. Starting point of our investigation is the quantum master 
equation in Lindblad form in Liouville space \cite{Ban93}. The concept of 
Liouville space to describe the evolution of the density matrix is well known 
and allows for a treatment similar to the Schr\"odinger equation in Hermitian 
quantum physics, with the difference that the Hamiltonian is replaced by a 
Liouvillian as the generator of the quantum master equation. The resulting time 
evolution is found by studying the induced Lie algebraic structure of this 
Liouvillian, either by diagonalisation of its regular representation, or by a 
Wei-Norman expansion.

The article is organised as follows. We begin our investigation with a 
discussion of the Liouville-space formalism in Sec.~\ref{sec:Liouville} where 
we introduce the system under study and formulate its generating Liouvillian. 
In Sec.~\ref{sec:LieAlgebra} we discuss how Lie algebraic structures emerge 
from this Liouvillian and prepare the ground to establish the methods for 
solving the quantum dynamics. Section~\ref{sec:eigendecomposition} 
focusses on solving the quantum dynamics for fixed system parameters by means of 
an eigendecomposition of the Liouvillian. The second method is a Wei-Norman 
expansion which is the subject of Sec.~\ref{sec:WeiNorman} which leads to 
deeper insight into the underlying algebraic structure. Concluding remarks can 
be found in Sec.~\ref{sec:Conclusion}. Some details regarding Liouville-space 
ladder operators and structure analysis in the Wei-Norman expansion have been 
relegated to the Appendix.
 
%%%%%%%%%%%%%%%%%%%%%%%%%%%%%%%%%%%%%%%%%%%%%%%%%%%%%%%%%%%%%%%%%%%%%%
\section{Liouville space formalism}
\label{sec:Liouville}

We begin our investigation by introducing the Liouville-space formalism 
\cite{Ban93} and show how to apply it to a system of coupled harmonic 
oscillators experiencing Markovian loss. A Liouville space $\mathfrak{L}$ is 
the Cartesian product of two Hilbert spaces. When used to describe open quantum 
systems, the involved spaces are usually the Hilbert space 
$\mathcal{H}$ of a closed system and its dual $\mathcal{H}'$, i.e. 
$\mathfrak{L} = \mathcal{H}\otimes \mathcal{H}'$. This effectively elevates the 
description from the quantum states $|\psi\rangle\in\mathcal{H}$ to the density 
matrices $\hat{\rho} \in \mathfrak{L}$. Operators in the Hilbert space $\hat{A} 
\in \mathcal{H}$ then become vectors in $\mathfrak{L}$ and are represented by
double kets $|A\rangle \rangle$. For these vectors, one can define a new inner 
product $\langle \langle A | B \rangle \rangle = \mathrm{Tr}  
\left( \hat{A}^\dagger \hat{B} \right)$. In conjunction with this inner product,
the Liouville space is itself a Hilbert space allowing many properties to 
be carried over to $\mathfrak{L}$, in particular notions such as projections 
and completeness. However, for this to be true some detailed attention has to 
be paid to the dual elements $\langle\langle A |$. The reason is that, in 
general, the dual space does not equal the Liouville space itself which is 
especially true for non-Hermitian open systems. This problem can be treated 
rigorously in the framework of rigged Hilbert spaces \cite{Gelfand}, and it can 
be shown \cite{Honda} that the Liouville space constructed from the Hilbert 
space spanned by bosonic Fock states does possess a complete basis.

When transitioning from the original Hilbert space to its derived Liouville 
space, the quantum dynamics is no longer given by the Schr\"odinger equation 
but by the von-Neumann equation, i.e.
\begin{align}
\label{eq:SchroedingerToVonNeumann}
\mi \frac{\mathrm{d}}{\mathrm{d}t} | \psi \rangle = \hat{H} | \psi \rangle 
\qquad \rightarrow \qquad \frac{\mathrm{d}}{\mathrm{d}t} | \rho \rangle \rangle 
= \mathcal{L} | \rho \rangle\rangle.
\end{align}
Here we introduced the Liouvillian $\mathcal{L}$ as a superoperator acting 
solely to the right. In order to define such a right action, superoperators can 
often be defined as left or right applications of Hilbert-space operators. For 
example, one can derive the two Liouville space superoperators $L[\hat{O}]$ and 
$R[\hat{O}]$ from the Hilbert space operator $\hat{O}$ as
\begin{align*}
 L[\hat{O}] | A \rangle \rangle = \hat{O} \hat{A},\\
 R[\hat{O}] | A \rangle \rangle = \hat{A} \hat{O}.
\end{align*}

This principle shall now be applied to a linear chain of $N$ coupled harmonic 
oscillators. The closed system Hamiltonian is 
\begin{equation}
\label{eq:H}
\hat{H} = \sum_{k=1}^N \sigma_k \hat{a}_k^\dagger \hat{a}_k + \sum_{k=1}^{N-1} 
\kappa_k \left( \hat{a}_k^\dagger \hat{a}_{k+1} + \hat{a}_k 
\hat{a}_{k+1}^\dagger\right)
\end{equation}
with the bosonic mode operators $\hat{a}_k$, $\hat{a}_k^\dagger$, the energy 
$\sigma_k$ of the $k$th mode, and the coupling $\kappa_k$ between modes $k$ and 
$k+1$. Such a Hamiltonian is often encountered, for example, when modelling 
light propagation in photonic waveguides \cite{Meany,Linares}.

Loss in the $k$th mode is introduced to the system as Markovian loss of a 
single excitation with rate $\gamma_k$ which, e.g., in photonic waveguides is 
due to scattering \cite{Guo81}. This is modelled by Lindblad terms in the von 
Neumann equation which then becomes the quantum master equation
\begin{equation}
\label{eq:MasterEqHilbert}
\frac{\mathrm{d}}{\mathrm{d}t} \hat{\rho} = - \mi \left[ \hat{H} , \hat{\rho} 
\right] + \sum_{k=1}^N \gamma_k \left( 2 \, \hat{a}_k \hat{\rho} \hat{a}_k^\dagger 
- \left\{ \hat{a}_k^\dagger \hat{a}_k, \hat{\rho} \right\} \right),
\end{equation}
where we used $\{ \bullet , \bullet\}$ to denote the anticommutator.

An interesting reformulation of the quantum master equation can be found by 
grouping the anticommutator term containing the Lindblad operators together
with the commutator containing the Hamiltonian of the closed system by defining
an effective non-Hermitian Hamiltonian $\hat{H}_\text{eff}$,
\begin{equation}
- \mi \left[ \hat{H} , \hat{\rho} \right] - \sum_{k=1}^N \gamma_k\left\{ 
\hat{a}_k^\dagger \hat{a}_k, \hat{\rho} \right\} = -\mi \left( 
\hat{H}_\text{eff} \hat{\rho} - \hat{\rho} \hat{H}^\dagger_\text{eff} \right),
\end{equation}
where we added the loss rate to the propagation constant, i.e. $\sigma_k 
\rightarrow \sigma_k - \mi \gamma_k$. Such a term would occur if one would
derive the von Neumann equation directly from the Schr\"odinger equation 
generated by the non-Hermitian $\hat{H}_\text{eff}$.
However, without the trace-preserving terms $2 \, \gamma_k \hat{a}_k \hat{\rho} 
\hat{a}_k^\dagger$ this effective Hamiltonian does not generate a physical 
evolution for the whole quantum state by itself. Nonetheless, in 
Sec.~\ref{sec:WeiNorman} we will see that it still plays an important role.

Using the bosonic mode operators we can now define a suitable set of 
superoperators 
\begin{gather}
L_k^- |A \rangle \rangle = \hat{a}_k \hat{A}, \qquad L_k^+ |A \rangle \rangle = 
\hat{a}_k^\dagger \hat{A}, \\
R_k^- |A \rangle \rangle = \hat{A} \hat{a}_k^\dagger, \qquad R_k^+ |A \rangle 
\rangle = \hat{A} \hat{a}_k.
\end{gather}
With these, the Liouvillian can solely be defined as a right-action superoperator
\begin{gather}
\mathcal{L} =\sum\limits_{k=1}^N \Big[ \left(\mi \sigma_k -\gamma_k\right) 
R_k^+R_k^- -\left(\mi \sigma_k+\gamma_k\right) L_k^+L_k^- \nonumber\\ 
+ 2 \gamma_k L_k^- R_k^- \Big]
 \nonumber\\
-\sum\limits_{k=1}^{N-1} \mi \kappa_k \left( L_k^+ L_{k+1}^- + L_{k+1}^+ 
L_k^- - R_{k+1}^+ R_k^- - R_k^+ R_{k+1}^- \right) .
\label{eq:L}
\end{gather}
Now that an explicit formulation of a right action is known we can also define 
a time-evolution superoperator $\mathcal{U}$ in Liouville space that yields the 
formal solution $|\rho(t)\rangle\rangle=\mathcal{U}(t)|\rho_0\rangle\rangle$.
This time-evolution superoperator $\mathcal{U}$ obeys the differential equation
\begin{equation}
 \label{eq:U_ODE}
\frac{\mathrm{d}}{\mathrm{d}t} \mathcal{U}(t)  = \mathcal{L} \, \mathcal{U}(t), 
\qquad \mathcal{U}(0) = \bm{1}_\mathcal{L}.
\end{equation}

In the following, we will discuss two approaches to solve the quantum master 
equation \eqref{eq:MasterEqHilbert} based on Lie algebras induced by 
Eq.~\eqref{eq:L}. The first approach is to find an eigendecomposition of a 
constant Liouvillian $\mathcal{L} \neq \mathcal{L}(t)$ that allows us to find 
an explicit form of the formal solution $\mathcal{U} = \me^{t \mathcal{L}}$. 
The second approach is applicable to a time-dependent Liouvillian 
$\mathcal{L}(t)$ and is based on a Wei-Norman expansion of $\mathcal{U}$. Both 
approaches are based on different Lie algebras induced by $\mathcal{L}$. Thus,
we first discuss how to generate those algebras from the explicit representation 
as seen in Eq.~\eqref{eq:L}.

%%%%%%%%%%%%%%%%%%%%%%%%%%%%%%%%%%%%%%%%%%%%%%%%%%%%%%%%%%%%%%%%%%%%%%
\section{Induced Lie algebras}
\label{sec:LieAlgebra}

From the considerations in the last section we can deduce that the linear 
chain of $N$ lossy harmonic oscillators are described by $2 N$ Hilbert-space operators 
$\{\hat{a}_i,\hat{a}_i^{\dagger}\}$ that result in $4 N$ Liouville-space 
operators $\{ L_i^\pm, R_i^\pm \}$. Those superoperators inherit their 
commutation relations from the canonical commutator 
$[\hat{a}_i,\hat{a}_j^{\dagger}] = \delta_{ij}$ of the Hilbert-space ladder 
operators. Explicitly, we find
\begin{equation}
 [L_i^-,L_j^+] = \delta_{ij}, \quad
 [R_i^-,R_j^+] = \delta_{ij}.
\end{equation}
Based on these fundamental commutators one can construct a Lie algebra spanned 
by all linear operators $\{ L_i^\pm, R_i^\pm \}$, all bilinear or quadratic 
operators $\{L_i^+ L_j^-, R_i^+ R_j^-, L_i^- R_j^- \}$, and the Liouville-space 
identity $\bm{1}_{\mathfrak{L}}$. Note that we included all mode combinations 
$i,j$ which derives from the fact that commutators such as $[L_i^+ L_{i+1}^-, 
L_{i+1}^+L_{i+2}^-] = L_i^+L_{i+2}^-$ occur. Hence, for the algebra to be closed 
we have to consider not only the nearest-neighbour coupling operators such as
$L_i^+ L_{i+1}^-$ that already occur in $\mathcal{L}$, but all other 
combinations as well. For this reason, the chosen linear-chain arrangement of 
the oscillators is only for ease of calculations and does not hinder the 
application to more complicated mode-coupling schemes.
Also note that one only needs to include the loss operators $L_i^- R_j^-$ as 
gain does not occur.

A closer inspection of the structure constants of the algebra spanned by 
$\{\bm{1}_{\mathfrak{L}},  L_i^\pm, R_i^\pm ,L_i^+ L_j^-, R_i^+ R_j^-, L_i^- 
R_j^- \}$ reveals that the linear operators, together with the identity,
constitute an ideal of the whole algebra. Thus, they can be separated to form 
their own subalgebra.
When including the Liouvillian $\mathcal{L}$ to this subalgebra we obtain a closed 
algebra spanned by $\{\bm{1}_{\mathfrak{L}},  L_i^\pm, R_i^\pm, \mathcal{L}\}$
that can be used to find an eigendecomposition of $\mathcal{L}$. 
Likewise, as $\mathcal{L}$ can be represented solely by the quadratic 
operators, see Eq.~\eqref{eq:L}, we can construct a closed subalgebra from them 
that leads to a Wei-Norman expansion of $\mathcal{L}$. Both closed algebras 
provide us with deeper insight into the lossy oscillator system and yield 
methods to solve the quantum master equation.

%%%%%%%%%%%%%%%%%%%%%%%%%%%%%%%%%%%%%%%%%%%%%%%%%%%%%%%%%%%%%%%%%%%%%%
\section{Eigendecomposition of the Liouvillian}
\label{sec:eigendecomposition}

The first method solves the quantum master equation for a time-independent 
Liouvillian $\mathcal{L}$ by means of a diagonalisation of its regular 
representation $\mathcal{R}(\mathcal{L})$. The regular representation of an 
element $Z$ of a Lie algebra spanned by $\{ X_i \}$ is defined as
\begin{equation}
\label{eq:defRegRepr}
 \mathcal{R}(Z): [Z,X_i] = \mathcal{R}_{ij}(Z) X_j
\end{equation}
and is thus intricately linked to its inherent structure constants. In case of 
the algebra spanned by $\{\bm{1}_{\mathfrak{L}},L_i^\pm,R_i^\pm,\mathcal{L}\}$, 
$\mathcal{R}(Z)$ is a $(4 N + 2) \times (4 N + 2)$-matrix. However, as we 
are only interested in $\mathcal{R}(\mathcal{L})$, we can drop the identity and 
$\mathcal{L}$ itself and focus on the smaller matrix $\mathcal{R}'(\mathcal{L})$
with elements $X_i$ drawn only from the linear operators $\{L_i^\pm, R_i^\pm\}$.
When arranging 
those linear operators as $\{L_1^+,\dots, L_N^+, R_1^-, \dots, R_N^-, R_1^+, 
\dots, R_N^+,L_1^-, \dots, L_N^-\}$, the regular representation becomes
\begin{equation}
\mathcal{R}'(\mathcal{L}) = \begin{pmatrix}
-\mi \bm{H}_\text{eff} & \bm{\Gamma} & \bm{0} & \bm{0} \\
\bm{0} & -\mi \bm{H}^\dagger_\text{eff} & \bm{0} & \bm{0} \\
\bm{0} & \bm{0} & \mi \bm{H}^\dagger_\text{eff} & \bm{\Gamma} \\
\bm{0} & \bm{0} & \bm{0} & \mi \bm{H}_\text{eff} 
\end{pmatrix}
\end{equation}
with $\bm{\Gamma} = 2 \,\mathrm{diag} (\gamma_1, \dots, \gamma_N)$ and
\begin{equation}
\bm{H}_\text{eff} = \begin{pmatrix}
\sigma_1 - \mi \gamma_1 & \kappa_1 & & \\
\kappa_1 & \ddots  & \ddots &\\
& \ddots & \ddots & \kappa_{N-1}\\
& &   \kappa_{N-1} &  \sigma_{N} - \mi \gamma_{N}
\end{pmatrix}.
\end{equation}

Diagonalisation of $\mathcal{R}'(\mathcal{L})$ yields $4 N$ eigenvalues
\begin{equation}
\label{eq:eigvalL}
\{\lambda_1, \dots, \lambda_N, -\lambda_1^*, \dots, -\lambda_N^*, \lambda_1^*, 
\dots, \lambda_N^*, -\lambda_1, \dots, -\lambda_N\},
\end{equation}
where $\lambda_i$ are the eigenvalues of $-\mi \bm{H}_\text{eff}$.
Their accompanying eigenvectors define $4 N$ superoperators that are 
superpositions of the linear operators $L_i^\pm, R_i^\pm$, and which we split 
into two equally large groups denoted by $P_i^\pm$ and $Q_i^\pm$.
They have the general form
\begin{gather}
 P^+_i = \sum_k c_{ik} (L^+_k - R^-_k),\quad
 P^-_i = \sum_k c_{ik} L^-_k,\\
 Q^+_i = \sum_k c^*_{ik} (R^+_k - L^-_k),\quad
 Q^-_i = \sum_k c^*_{ik} R^-_k. 
\end{gather}
These new superoperators act as collective creation and annihiliation 
operators in Liouville space. In fact, after suitable normalisation, they obey 
the commutator relations
\begin{equation}
 [P_i^-,P_j^+] = \delta_{ij}, \quad
 [Q_i^-,Q_j^+] = \delta_{ij},
\end{equation}
with all other commutators vanishing. Furthermore, because they diagonalise the 
regular representation $\mathcal{R}'(\mathcal{L})$, they yield the eigenvalue 
commutator relations
\begin{gather}
 \left[ \mathcal{L}, P_i^+ \right] = \lambda_i P_i^+,\quad
 \left[ \mathcal{L}, P_i^- \right] = -\lambda_i P_i^-,\nonumber\\
 \left[ \mathcal{L}, Q_i^+ \right] = \lambda_i^* Q_i^+,\quad
 \left[ \mathcal{L}, Q_i^- \right] = -\lambda_i^* Q_i^-.
\label{eq:commutators}
\end{gather}
These commutators are reminiscent of the ladder operators in Hilbert space 
which is the reason why we termed them creation and annihiliation operators 
that add/subtract collective excitations of $\pm\lambda_i$ or $\pm\lambda_i^*$ 
in Liouville space. Note, however, that these are not physical excitations but 
are only to be understood as abstract excitations in $\mathfrak{L}$ because 
they contain creation and annihiliation operators from the original Hilbert 
space $\mathcal{H}$.

Nonetheless, the Liouville space ladder operators can be used to construct 
eigenvectors of the Liouvillian $\mathcal{L}$. First, we define a ground state 
$|0\rangle\rangle$ via
\begin{gather}
 \label{eq:rightGround}
 P_i^-|0 \rangle \rangle = Q_i^-|0\rangle \rangle = 0, \qquad \forall i,\\
 \label{eq:leftGround}
 \langle \langle 0| P_i^+ =\langle \langle 0| Q_i^+ = 0, \qquad \forall i,
\end{gather}
where one has to keep in mind that, due to the non-Hermiticity of $\mathcal{L}$,
the left and right vectors are different. A careful analysis leads to
\begin{equation}
 |0 \rangle \rangle = |\bm{0} \rangle \langle \bm{0}| ,\quad
 \langle \langle 0| = \bm{1}_\mathcal{H}
\end{equation}
with $\bm{1}_\mathcal{H}$ as the identity in the base Hilbert space, see 
Appendix~\ref{appdx:ladder}. Higher rungs of the ladder are then obtained as
\begin{align}
|\bm{\alpha},\bm{\beta} \rangle\rangle &= 
\frac{1}{\sqrt{\bm{\alpha}!\bm{\beta}!}} \bm{P}^{+ \,\bm{\alpha}} \bm{Q}^{+ 
\,\bm{\beta}} |0\rangle\rangle\\
\label{eq:eigvecConstr}
\langle\langle \bm{\alpha},\bm{\beta} | &= 
\frac{1}{\sqrt{\bm{\alpha}!\bm{\beta}!}} \langle\langle 0| \bm{P}^{- 
\,\bm{\alpha}} \bm{Q}^{- \,\bm{\beta}},
\end{align}
with multiindeces $\bm{\alpha} = (\alpha_1, \dots,\alpha_{N})$.

The Liouville-space ladder states are eigenstates of the Liouvillian 
$\mathcal{L}$ which can be seen by using the commutators \eqref{eq:commutators} 
and the fact that $\mathcal{L}|0 \rangle \rangle = 0$. Similarly, we can 
calculate the action of the evolution operator $\mathcal{U}=\me^{t\mathcal{L}}$ 
on the states $|\bm{\alpha}, \bm{\beta}\rangle\rangle$ via the adjoint action 
defined by
\begin{equation}
 \me^{A} B \me^{-A} = B + [A,B] + \frac{1}{2} [A,[A,B]] + \dots
\end{equation}
Applying this to the product of creation operators yields after straightforward 
calculation
\begin{gather}
\me^{t \mathcal{L}} \bm{P}^{+ \,\bm{\alpha}} \bm{Q}^{+ \,\bm{\beta}} 
= \me^{t\mathcal{L}} \bm{P}^{+ \,\bm{\alpha}} 
\bm{Q}^{+\,\bm{\beta}}\me^{-t\mathcal{L}} \me^{t \mathcal{L}} \nonumber\\
=\me^{t \,( \bm{\alpha} \cdot \bm{\lambda} + \bm{\beta} \cdot \bm{\lambda}^*)} 
\bm{P}^{+ \,\bm{\alpha}} \bm{Q}^{+ \,\bm{\beta}} \me^{t \mathcal{L}},
\end{gather}
with $\bm{\lambda} = (\lambda_1, \dots, \lambda_N)$. The action on a 
ladder state thus becomes
\begin{equation}
\me^{t \mathcal{L}}  |\bm{\alpha}, \bm{\beta} \rangle\rangle = \me^{t \,( 
\bm{\alpha} \cdot \bm{\lambda} + \bm{\beta} \cdot \bm{\lambda}^*)}  
|\bm{\alpha}, \bm{\beta} \rangle\rangle.
\end{equation}
With this result, the time-evolved quantum state is
\begin{equation}
\label{eq:EigDecompSol}
|\rho (t) \rangle \rangle = \me^{t \mathcal{L}} |\rho_0 \rangle \rangle = 
\sum_{\bm{\alpha}, \bm{\beta}} \me^{t \,( \bm{\alpha} \cdot \bm{\lambda} + 
\bm{\beta} \cdot \bm{\lambda}^*)}  |\bm{\alpha}, \bm{\beta} \rangle\rangle 
\langle \langle \bm{\alpha}, \bm{\beta} | \rho_0 \rangle \rangle.
\end{equation}
The sum runs over all possible multiindices which, however, in practice are 
limited to a select few. In fact, for an initial state $\hat{\rho}_0$ 
containing $N_\text{p}$ bosonic excitations, the multiindices are restricted by 
$|\bm{\alpha}|,|\bm{\beta}|\leq N_\text{p}$.

The expression Eq.~\eqref{eq:EigDecompSol} is an analytic solution to the 
quantum master equation \eqref{eq:MasterEqHilbert} of a fixed system of coupled 
lossy harmonic oscillators. For a few modes and excitations, explicit results 
can be calculated by hand. For example, let us consider the case of two 
harmonic oscillators ($N=2$). We define the differences of the propagation 
constants $\sigma_1$ and $\sigma_2$ and loss rates $\gamma_1$ and $\gamma_2$ as 
well as their respective means as
\begin{gather}
\Delta \sigma = \frac{\sigma_1 - \sigma_2}{2}, \qquad \overline{\sigma} = 
\frac{\sigma_1 + \sigma_2}{2}, \\
\Delta \gamma = \frac{\gamma_1 - \gamma_2}{2}, \qquad \overline{\gamma} = 
\frac{\gamma_1 + \gamma_2}{2}.
\end{gather}
Together with the (single) coupling rate $\kappa_1 =\kappa$ we find for the two 
eigenvalues of $-\mi \bm{H}_\text{eff}$
\begin{equation}
\label{eq:eigvalHeff}
 \lambda_1 = - \overline{\gamma} - \mi \overline{\sigma} + \mi \omega,\quad
 \lambda_2 = - \overline{\gamma} - \mi \overline{\sigma} - \mi \omega,
\end{equation}
with
\begin{equation}
 \omega = \sqrt{\kappa^2 + (\Delta \sigma - \mi \Delta \gamma)^2}.
\end{equation}
From these, the eight eigenvalues of the Liouvillian are constructed as in 
Eq.~\eqref{eq:eigvalL}. The diagonalisation of the regular representation 
$\mathcal{R}' (\mathcal{L})$ yields two sets of bosonic ladder superoperators, 
the first being
\begin{gather}
P_i^+ = \epsilon_i \left( L_1^+ - R_1^- \right) + \tau_i \left( L_2^+ - R_2^- 
\right),\\
P_i^- = \epsilon_i L_1^- + \tau_i L_2^-
\end{gather}
and the second
\begin{gather}
Q_i^+ = \epsilon_i^* \left( R_1^+ - L_1^- \right) + \tau_i^* \left( R_2^+ - 
L_2^- \right),\\
Q_i^- = \epsilon_i^* R_1^- + \tau_i^* R_2^-
\end{gather}
with
\begin{gather}
 \epsilon_1 = \frac{-\kappa}{\sqrt{2 \omega (\omega + \Delta \sigma - \mi \Delta 
\gamma)}}, \;\; 
 \tau_1 = \frac{\omega + \Delta \sigma - \mi \Delta \gamma}{\sqrt{2 \omega 
(\omega + \Delta \sigma - \mi \Delta \gamma)}},\\
 \epsilon_2 = \frac{ \kappa}{\sqrt{2 \omega (\omega - \Delta \sigma + \mi \Delta 
\gamma)}}, \;\;
 \tau_2 = \frac{\omega - \Delta \sigma + \mi \Delta \gamma}{\sqrt{2 \omega 
(\omega - \Delta \sigma + \mi \Delta \gamma)}}.
\end{gather}

With the ladder superoperators at hand, the overlap $\langle\langle\bm{\alpha}, 
\bm{\beta}|\rho_0\rangle\rangle$ from Eq.~\eqref{eq:EigDecompSol} can be 
straightforwardly calculated using the definition 
of the Liouville space inner product and the construction rule for the left 
eigenvectors of $\mathcal{L}$ in Eq. \eqref{eq:eigvecConstr} as
\begin{equation}
\langle \langle \bm{\alpha}, \bm{\beta} | \rho_0 \rangle \rangle 
=\frac{\mathrm{Tr} \left(  P_1^{- \,\alpha_1} P_2^{- \,\alpha_2} Q_1^{- 
\,\beta_1} Q_2^{- \,\beta_2} 
\hat{\rho}_0\right)}{\sqrt{\alpha_1!\beta_1!\alpha_2!\beta_2!}}.
\end{equation}
This expression depends on the input state $\hat{\rho}_0$ and determines which 
eigenvectors $|\bm{\alpha}, \bm{\beta} \rangle\rangle$ have to be considered.
Together with the exponential function involving the eigenvalues $\lambda_i$ 
and $\lambda_i^*$ one can then calculate the time-evolved state and thus any 
observable.

This method is well suited to theoretically describe the experiment on passive 
$\mathcal{PT}$-symmetric waveguide couplers \cite{Klauck19}. The coupler 
consisted of two waveguides at a fixed spatial separation, and hence with a 
fixed coupling strength, where in one of them a loss rate $\gamma$ 
was introduced by periodic bending of the waveguide. In the experiments,
the effect of the loss, and with that of the increasing non-Hermiticity, on 
the coincidence rate $\Gamma = \langle\hat{a}_1^\dagger\hat{a}_2^\dagger 
\hat{a}_1\hat{a}_2 \rangle$ for detecting two photons was measured.
The two photons were initially launched in each waveguide separately, i.e.
the initial state was $|1,1\rangle$.
The analytical solution for $\Gamma$ can be easily calculated using the above 
method and reads
\begin{align}
\label{eq:PTHOM}
\Gamma = \mathrm{e}^{- 2 \gamma t} \left| \frac{\gamma^2 - 4 \kappa^2 \cos 
\left( t \sqrt{4 \kappa^2 - \gamma^2} \right)}{4 \kappa^2 - \gamma^2} \right|^2.
\end{align}

\begin{figure}
 \includegraphics[width=\columnwidth]{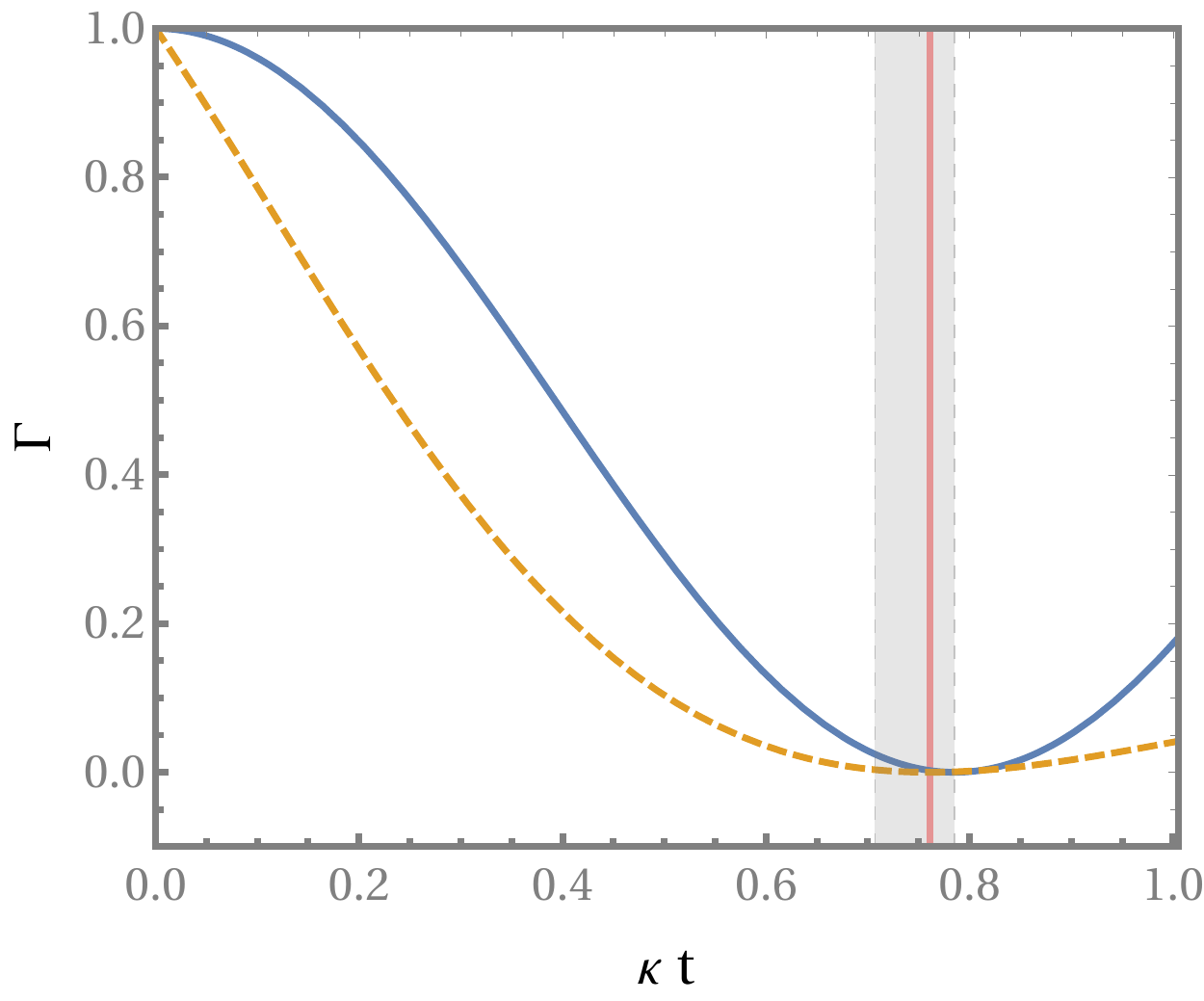}
\caption{The coincidence function \eqref{eq:PTHOM} of two coupled harmonic 
oscillators, one of which experiences loss $\gamma$, with an initial excitation 
$|1,1\rangle$. Blue solid line: Hermitian case $\gamma = 0$. Orange dashed line:
$\mathcal{PT}$-symmetric case with $\gamma = \kappa$, the red vertical line 
indicates the position of the shifted bunching ($\Gamma = 0$). The shaded area 
shows the range of the shift for which the system is still $\mathcal{PT}$ 
symmetric, i.e. $\gamma \in [0,2 \kappa]$. 
% This result is the theory behind the measurements on waveguides in 
% Ref.~\cite{Klauck19}.
}
\label{fig:coinExp}
\end{figure}

In Fig.~\ref{fig:coinExp}, the coincidence rate for two values of the loss 
$\gamma$ is plotted against the normalised time $\kappa t$. In the Hermitian 
case $\gamma = 0$ (blue line) the well-known result from the Hong-Ou-Mandel 
experiment is reproduced. The photons bunch at $\kappa t = \pi/4$ and leave the 
system together in either of the two modes. With increased non-Hermiticity, 
however, the bunching (red vertical line) is shifted to shorter values of 
$\kappa t$ with one example with $\gamma = \kappa$ shown (orange dashed). 
Note that the $\mathcal{PT}$ phase is limited to $\gamma \leq 2 \kappa$, the 
grey area shows the maximal extend of the shift down to $\kappa t = 1/\sqrt{2}$.
In Ref.~\cite{Klauck19}, these prediction where experimentally confirmed.

Before continuing, we would like to add a few remarks. Because the solution of 
the quantum master equation in Eq.~\eqref{eq:EigDecompSol} is based on an 
eigendecomposition of the Liouvillian, we can analyse the evolution of the 
lossy system by first examining the eigenvalues $( \bm{\alpha}\cdot\bm{\lambda} 
+\bm{\beta}\cdot\bm{\lambda}^*)$. Clearly, certain combinations of 
multiindices $\bm{\alpha}$ and $\bm{\beta}$ will result in lower total loss 
rates than others. In this way, one can carefully 
design the lossy system and/or choose the ideal input states that are 
(predominantly) expanded in Liouville eigenstates $|\bm{\alpha}, \bm{\beta} 
\rangle\rangle$ to minimise the detrimental effects of the losses. For example,
in the case $N=2$ with eigenvalues as given in Eq.~\eqref{eq:eigvalHeff}, upon
choosing $\Delta \sigma = 0$ and $\Delta \gamma > \kappa$ we find
eigenvalues with $\lambda_{1,2} = \overline{\gamma} \pm \delta \in \mathbb{R}$ 
and thus the appearance of states that propagate with loss lower than the mean 
$\overline{\gamma}$. This is the so-called $\mathcal{PT}$-broken 
phase that exists beyond the exceptional point $\Delta \gamma = \kappa$.

Furthermore, our approach based on an eigendecomposition of the Liouvillian can 
be extended to analytically solve the quantum dynamics at exceptional points 
itself. For that, one has to construct the Jordan decomposition of the regular 
representation using standard techniques to construct the missing eigenvectors 
of the defective matrix $\mathcal{R}'(\mathcal{L})$.

%%%%%%%%%%%%%%%%%%%%%%%%%%%%%%%%%%%%%%%%%%%%%%%%%%%%%%%%%%%%%%%%%%%%%%
\section{Wei-Norman Expansion}
\label{sec:WeiNorman}

The Wei-Norman expansion \cite{WeiNorman63} is a method to solve first-order 
differential equations such as Eq.~\eqref{eq:U_ODE} when the generator, in our 
case the Liouvillian, has the form
\begin{align*}
 \mathcal{L}(t) = \sum_{k=1}^m c_k (t) X_k,
\end{align*}
where the $X_k$ constitute a set of constant operators. This set of operators 
induces a Lie algebra which can be closed under commutation by addition of 
suitable operators resulting in a set $\{ X_1, \dots , X_n \}$ with $n \geq m$.
With this prerequisite fulfilled, the Wei-Norman expansion is a product of 
individual exponentials
\begin{equation}
 \label{eq:WeiNormanExpansion}
\mathcal{U}(t) = \prod_{k=1}^n \mathcal{U}_k(t) = \prod_{k=1}^n \exp \left[
g_k(t) X_k \right]
\end{equation}
where each factor under the product solves its own differential equation
\begin{align*}
\frac{\mathrm{d}}{\mathrm{d}t} \mathcal{U}_k (t) = \dot{g}_k (t)  X_k 
\mathcal{U}_k(t), \qquad \mathcal{U}_k(0) = \bm{1}_\mathcal{L}.
\end{align*}
The functions $g_k(t)$ are solutions to a set of nonlinear differential 
equations that can be derived by differentiating the ansatz 
\eqref{eq:WeiNormanExpansion} and comparing it to the original differential 
equation \eqref{eq:U_ODE}.

The set of nonlinear differential equations for $g_k(t)$ is always integrable 
if the Lie algebra spanned by $\{X_1, \dots, X_n\}$ is solvable. If the Lie 
algebra is not solvable, we can, however, decompose it into a solvable part (its 
radical) and a remaining semisimple subalgebra \cite{WeiNorman63}. This means 
that the generator is split as $\mathcal{L} = \mathcal{L}_\text{R} + 
\mathcal{L}_\text{S}$, and the time-evolution operator becomes a product 
$\mathcal{U} = \mathcal{U}_\text{S} \mathcal{U}_\text{R}$ where each part obeys
a separate differential equation
\begin{gather}
\frac{\mathrm{d}}{\mathrm{d} t} \mathcal{U}_\text{S} = \mathcal{L}_\text{S} 
\mathcal{U}_\text{S},\nonumber\\
\frac{\mathrm{d}}{\mathrm{d} t} \mathcal{U}_\text{R} = \left( 
\mathcal{U}^{-1}_\text{S} \mathcal{L}_\text{R} \mathcal{U}_\text{S} \right) 
\mathcal{U}_\text{R}.
\label{eq:US}
\end{gather}
The solvable part $\mathcal{U}_\text{R}$ is easily integrated once the 
semisimple part is solved and $\mathcal{U}^{-1}_\text{S} \mathcal{L}_\text{R} 
\mathcal{U}_\text{S}$ is calculated. Thus, the actual difficulty is usually to 
find the solution for the semisimple subalgebra which can be broken down using 
the structure theorem for the decomposition of semisimple Lie algebras into a 
direct sum of simple Lie algebras. Although an analytical solution cannot be 
found in all cases, it turns out that, in our particular case, we will find 
solutions (at least numerically).

From the above outline of the Wei-Norman expansion one observes that the 
important task is the analysis of the Lie algebraic structure induced by 
$\mathcal{L}$. In case of the lossy oscillator system, the operators $X_k$ are 
drawn from the set $\{ L_i^+ L_j^-, R_i^+ R_j^-, L_i^- R_j^- \}$ as discussed 
in Sec.~\ref{sec:LieAlgebra}. These operators can be decomposed as follows:
\begin{gather}
\underbrace{\overbrace{\{ L_i^- R_j^- \} }^{\text{nilpotent}} \oplus 
\overbrace{\{\sum_k L_k^+L_k^- , \sum_k 
R_k^+R_k^-\}}^{\text{Abelian}}}_{\text{solvable (radical)}} \nonumber\\ 
\oplus \{ L_{k}^+L_{k}^- - L_{k+1}^+L_{k+1}^-, L_i^+L_{j\neq i}^- \} \nonumber\\
\oplus  \{ R_{k}^+R_{k}^- - R_{k+1}^+R_{k+1}^-, R_i^+R_{j\neq i}^-\} .
\label{eq:AlgebraStruct}
\end{gather}
The nilpotent subalgebra $\{ L_i^- R_j^- \}$ is to be understood as containing 
all combinations of indices $i,j=1,\dots,n$ and is responsible for removing 
excitations from the oscillators. The Abelian part contains the sum over all 
number operators, acting from the left and the right. This contribution is 
responsible for the mean loss of the oscillator system as befits a passive 
$\mathcal{PT}$-symmetric system.
% for exponential factors involving a mean complex energy term,
% and results in a separated factor. 
Together, both parts form the solvable subalgebra.

What is left are the difference operators, e.g. $ L_{k}^+L_{k}^- - L_{k+1}^+L_{k+1}^-$,
and the coupling operators $L_i^+L_{j\neq i}^-$. These form two special linear Lie algebras 
$\mathfrak{sl}(n,\mathbb{C})$, one for the left and one for the right 
application, that create the semisimple part. For more details of how to arrive 
at the given algebraic structure, we refer the reader to 
Appendix~\ref{appdx:AlgebraStruc}.

An immediate result of this structure analysis is that the excitation-removing operators
are separated from the rest of the dynamics in the Wei-Norman expansion.
This means that as long as the dynamics is projected onto the same Fock layer 
as the initial state, these parts do not contribute. For example, when the 
system is initialised with $N_\text{p}$ excitations and measurements are 
postselected to this number of excitations, the excitation-removing operators do 
not contribute, and the dynamics can be described by the effective 
non-Hermitian Hamiltonian $\hat{H}_\text{eff}$.

This fact can already be deduced from the quantum master equation 
\eqref{eq:MasterEqHilbert} where we already showed that the anticommutator term 
can be included in the Hamiltonian to create $\hat{H}_\text{eff}$.
The remaining part $\sum_k \gamma_k \hat{a}_k \hat{\rho}\hat{a}_k^\dagger$ of 
the Lindblad term removes bosonic excitations and hence cannot contribute to any 
postselected measurement in which all initial excitations remain in the system.
This result justifies the naive approach of using the effective 
non-Hermitian Hamiltonian $\hat{H}_\text{eff}$ when modelling lossy quantum 
systems as long as measurements are restricted to the highest Fock layer.

Such a post-selection condition that results in an effective non-Hermitian evolution
is closely related to the quantum jump method to unravel the quantum master 
equation \cite{Dalibard,Plenio}. This method stochastically evolves wave 
functions by repeatedly applying an effective non-Hermitian Hamiltonian over a 
time increment after which a quantum jump randomly may or may not occur. 
Averaging over a sufficiently large number of such evolutions (or quantum 
trajectories) yields the same dynamics as the quantum master equation. 
A post-selection of situations without quantum jumps is thus equivalent to a 
quantum Zeno dynamics\cite{Beige97,Itano90} which is solely determined by the non-Hermitian 
Hamiltonian.

As a final remark we note that the deduced structure also shows a separation
of the mean energy constant and mean losses, i.e. the Abelian part, from the rest of 
the Hamiltonian. This justifies the use of passive non-Hermitian systems, 
such as passive $\mathcal{PT}$-symmetric systems, because the mean loss can 
indeed be separated from the important dynamics induced by the special linear 
Lie algebra $\mathfrak{sl}(n,\mathbb{C})$. Thus, a non-Hermitian system with 
loss and gain can be simulated by a system with only loss after post-selection 
and correction of the overall loss (see the example in Eq.~\eqref{eq:PTHOM}), 
but without the added noise from the gain process \cite{Scheel}.

%%%%%%%%%%%%%%%%%%%%%%%%%%%%%%%%%%%%%%%%%%%%%%%%%%%%%%%%%%%%%%%%%%%%%%
\subsection{Example of two harmonic oscillators in Wei-Norman expansion}

In the following, we will showcase how to apply the Wei-Norman expansion by 
considering the simplest nontrivial case of two coupled oscillators.
The first step is to separate the Liouvillian into the radical and semisimple 
parts so that we can solve the equations \eqref{eq:US} for the individual 
time-evolution superoperators. Starting from the general Liouvillian 
\eqref{eq:L} and using the decomposition \eqref{eq:AlgebraStruct}, we find for 
the radical part
\begin{gather}
\mathcal{L}_\text{R} (t) = \frac{1}{2} \left( -\mathrm{i} (\sigma_1 + \sigma_2) 
- (\gamma_1 + \gamma_2) \right) \left( L_1^+L_1^- + L_2^+L_2^- \right) 
\nonumber\\
+ \frac{1}{2} \left( \mathrm{i} (\sigma_1 + \sigma_2) - (\gamma_1 + \gamma_2) 
\right) \left( R_1^+R_1^- + R_2^+R_2^- \right) \nonumber\\
+ 2\gamma_1 L_1^- R_1^- + 2\gamma_2 L_2^- R_2^-,
\end{gather}
where one clearly observes the emergence of mean values for the energy 
constants and the losses as discussed earlier. As for the semisimple part, we 
know that it is comprised of two isomorphic and commuting simple parts so that 
it can be split as $\mathcal{L}_\text{S} (t) = \mathcal{L}_\text{S$_1$} (t) 
+\mathcal{L}_\text{S$_2$} (t)$ with
\begin{gather}
\mathcal{L}_\text{S$_1$} (t) = \frac{1}{2} \left( -\mathrm{i} (\sigma_1 - 
\sigma_2) - (\gamma_1 - \gamma_2) \right) \left( L_1^+L_1^- - L_2^+L_2^- \right) 
\nonumber\\
- \mathrm{i} \kappa \left( L_1^+L_2^- + L_2^+L_1^- 
\right),\label{eq:ExplCalcLS1}
\\
\mathcal{L}_\text{S$_2$} (t) = \frac{1}{2} \left( \mathrm{i} (\sigma_1 - 
\sigma_2) - (\gamma_1 - \gamma_2) \right) \left( R_1^+R_1^- - R_2^+R_2^- \right) 
\nonumber\\
+ \mathrm{i} \kappa \left( R_1^+R_2^- + R_2^+R_1^- \right).
 \label{eq:ExplCalcLS2}
\end{gather}
Note that $\mathcal{L}_\text{S$_1$}$ and $\mathcal{L}_\text{S$_2$}$ transform 
into one another under exchange of left and right actions, and a complex 
conjugation of the possibly $t$-dependent prefactors. Both solve their own 
respective differential equations
\begin{equation}
 \label{eq:SimplePartODE}
\frac{\mathrm{d}}{\mathrm{d} t} \mathcal{U}_\text{S$_j$} = 
\mathcal{L}_\text{S$_j$} \mathcal{U}_\text{S$_j$}.
\end{equation}
Because both sets are isomorphic we can solve both using one operator representation.
In the present case this is the special linear algebra $\mathfrak{sl} (2, \mathbb{C})$
\begin{gather}
 K_0 = L_1^+L_1^- - L_2^+L_2^- \vee R_1^+R_1^- - R_2^+R_2^- ,\\
 K_+ = L_1^+L_2^- \vee R_1^+R_2^- ,\\
 K_- = L_2^+L_1^- \vee R_2^+R_1^-,
\end{gather}
with commutators
\begin{equation}
 [K_0 , K_\pm] = \pm 2 K_\pm ,\quad
 [K_+ , K_-] = K_0.
\end{equation}

Our ansatz for the Wei-Norman expansion is
\begin{equation}
\mathcal{U}_\text{S$_1$} (t) = \mathrm{e}^{f_+ (t) K_+} \mathrm{e}^{f_0 (t) K_0} 
\mathrm{e}^{f_- (t) K_-}
\end{equation}
and analagously for $\mathcal{U}_\text{S$_2$}$ using the respective complex 
conjugate functions. Inserting this ansatz into the differential equation 
\eqref{eq:SimplePartODE} and carefully calculating the required commutator 
relations \cite{Korsch}, we derive the set of nonlinear differential equations 
for the functions $f_\pm (t)$ and $f_0 (t)$ as
\begin{gather}
 \dot{f}_- \mathrm{e}^{- 2 f_0 } = - \mathrm{i} \kappa ,\\
 \dot{f}_0 + \dot{f}_- f_+ \mathrm{e}^{-2 f_0} = - \mathrm{i} \Delta ,\\
 \dot{f}_+ - 2 \dot{f}_0 f_+ - \dot{f}_- f_+^2 \mathrm{e}^{-2 f_0} = - 
\mathrm{i} \kappa,
\end{gather}
where we defined $-\frac{1}{2} \left[(\sigma_1 - \sigma_2) + \mi (\gamma_1 - 
\gamma_2) \right] = \Delta$.
This set of nonlinear differential equations can be reformulated 
as a Riccati differential equation for $f_+$, i.e.
\begin{equation}
 \dot{f}_+ + \mathrm{i} 2 \Delta f_+ - \mathrm{i} \kappa f_+^2 + \mathrm{i} \kappa = 0.
\end{equation}

Riccati equations are in principle solvable as they can be reduced to 
first-order differential equations. Once the solution is found, one can 
integrate the remaining equations
\begin{gather}
 \dot{f}_0 = - \mathrm{i} \Delta + \mathrm{i} \kappa f_+ ,\\
 \dot{f}_- = - \mathrm{i} \kappa \, \mathrm{e}^{2 f_0}.
\end{gather}
This gives the functions $f_\pm$, $f_0$ which solve the $\mathfrak{sl} 
(2,\mathbb{C})$ problem for the left superoperators. Due to the isomorphy of 
the right superoperators, their result differs only by a complex conjugation of 
the functions $f_\pm$, $f_0$. The total solution of the semisimple part is then 
given by the time-evolution superoperator 
$\mathcal{U}_\text{S}=\mathcal{U}_\text{S$_1$} \mathcal{U}_\text{S$_2$}$ with
\begin{gather}
\mathcal{U}_\text{S$_1$} = \mathrm{e}^{f_+ L_1^+L_2^-} \mathrm{e}^{f_0 \left( 
L_1^+L_1^- - L_2^+L_2^- \right)} \mathrm{e}^{f_- L_2^+L_1^-},\\
\mathcal{U}_\text{S$_2$} = \mathrm{e}^{f_+^* R_1^+R_2^-} \mathrm{e}^{f_0^* 
\left( R_1^+R_1^- - R_2^+R_2^- \right)} \mathrm{e}^{f_-^* R_2^+R_1^-}.
\label{eq:USsolution}
\end{gather}
This structure should not be a surprise as the right action of the 
superoperators simply results in the well-known time evolution 
$\mathcal{U}_\text{S} | \rho \rangle \rangle = U \rho U^\dagger$ where $U$ is 
the evolution operator determined by the effective Hamiltonian with the mean 
energy constant and mean loss removed.

Having solved the semisimple part of the algebra, the radical part is solved 
using Eq.~\eqref{eq:US}. Knowing the right-hand side $ \left( 
\mathcal{U}_\text{S}^{-1} 
\mathcal{L}_\text{R} \mathcal{U}_\text{S} \right) \mathcal{U}_\text{R}$ and 
using the ansatz
\begin{gather}
\mathcal{U}_\text{R} = \mathrm{e}^{a_1 (t) \left( L_1^+L_1^- + L_2^+L_2^- 
\right)} \mathrm{e}^{a_2 (t) \left( R_1^+R_1^- + R_2^+R_2^- \right)} \nonumber\\
\times\mathrm{e}^{a_3 (t) L_1^- R_1^-} \mathrm{e}^{a_4 (t) L_2^- R_2^-}
 \mathrm{e}^{a_5 (t) L_2^- R_1^-} \mathrm{e}^{a_6 (t) L_1^- R_2^-},
\label{eq:URsolution}
\end{gather}
the set of equations for the functions $a_i(t)$ with initial conditions $a_i(0) 
= 0$ can be calculated and reads
\begin{gather}
\dot{a}_1 = \frac{1}{2} \left[ - \mathrm{i} \left( \sigma_1 + \sigma_2 \right) 
- \left( \gamma_1 + \gamma_2 \right) \right] , \\
\dot{a}_2 = \frac{1}{2} \left[  \mathrm{i} ( \sigma_1 + \sigma_2 ) -  ( 
\gamma_1 + \gamma_2 ) \right] , \\
\dot{a}_3 = 2\mathrm{e}^{a_1 + a_2} \left( \gamma_1 \left| \mathrm{e}^{f_0} + 
\mathrm{e}^{-f_0} f_+ f_- \right|^2 + \gamma_2 \mathrm{e}^{-2 \mathrm{Re} 
(f_0)} |f_-|^2 \right) ,\\
\dot{a}_4 = 2\mathrm{e}^{a_1 + a_2}\left( \gamma_1 \mathrm{e}^{- 2\mathrm{Re} 
(f_0)} |f_+|^2 + \gamma_2 \mathrm{e}^{- 2 \mathrm{Re} (f_0)} \right) ,\\
\dot{a}_5 = 2\mathrm{e}^{a_1 + a_2}\bigg[ \gamma_1 \left( \mathrm{e}^{f_0^*} + 
\mathrm{e}^{-f_0^*} f_+^* f_-^* \right) \mathrm{e}^{-f_0} f_+ \nonumber\\
+ \gamma_2 \mathrm{e}^{- 2 \mathrm{Re} (f_0)} f_-^* \bigg] ,\\
\dot{a}_6 = 2\mathrm{e}^{a_1 + a_2}\bigg[ \gamma_1 \left( \mathrm{e}^{f_0} + 
\mathrm{e}^{-f_0} f_+ f_- \right) \mathrm{e}^{-f_0^*} f_+^* \nonumber\\
+ \gamma_2 \mathrm{e}^{- 2 \mathrm{Re} (f_0)} f_- \bigg].
\end{gather}
As expected, this set is uncoupled and thus directly integrable.
The functions $a_1(t)$ and $a_2(t)$ determine the Abelian contribution and are 
given by integrals of the (generally $t$-dependent) mean energy constant and 
and mean loss rates. All other functions $a_i(t)$ with $i=3,\dots,6$ determine the 
excitation-removing operations.

The solutions for the time-evolution superoperators of the semisimple part,
Eqs.~\eqref{eq:USsolution}, together with the solution for the radical part in 
Eq.~\eqref{eq:URsolution}, yield the total time evolution of a quantum state 
in the lossy waveguide system and is generally applicable for $t$-dependent 
system parameters. For example, the case of a $\mathcal{PT}$-coupler solved 
with the eigendecomposition in Sec.~\ref{sec:eigendecomposition} can now be 
generalised to $t$-dependent waveguides including nonzero propagation constants 
$\sigma_k$. For that, we start again with the input state $\hat{\rho}(0) = 
|1,1\rangle\langle 1,1|$ and calculate the coincidence rate resulting in
\begin{equation}
 \Gamma = \mathrm{e}^{2(a_1+a_2)} \left| 1 + 2 f_+ f_- \, \mathrm{e}^{-2 f_0}  
\right|^2.
\end{equation}
Note that only the functions $f_\pm$, $f_0$, and $a_{1,2}$ occur because the 
measurement is restricted to the Fock layer of two excitations and the 
excitation-removing operators of $\mathcal{U}_\text{R}$ in Eq.~\eqref{eq:URsolution}
do not contribute.

%%%%%%%%%%%%%%%%%%%%%%%%%%%%%%%%%%%%%%%%%%%%%%%%%%%%%%%%%%%%%%%%%%%%%%
\section{Conclusions}
\label{sec:Conclusion}

In this work, we discussed two methods for solving the quantum master equation 
of $N$ coupled bosonic modes that experience Markovian losses. Both methods 
work in a Liouville-space framework and rely on different aspects of the Lie 
algebra induced by the Liouvillian that generates the quantum master equation.
First, we diagonalised the regular representation of the Liouvillian to obtain 
its Liouville space eigenvectors and ladder superoperators. These allowed to 
evolve a quantum state for fixed system parameters. Additionally, they can be 
utilised to examine the transport properties of the system by analysing the 
eigenvalues of the Liouvillian.

Second, we employed a Wei-Norman expansion of the time-evolution operator that 
provides not only the solution for a time-varying system but also gives deeper 
insight into the underlying algebraic structure. The latter shows a clear 
separation of the dynamics induced by a non-Hermitian effective Hamiltonian 
from the excitation-removing operations which means that, measurements that are
postselected to outcomes in which no excitations have been lost, can solely be 
described by this effective Hamiltonian. Furthermore, from this effective 
Hamiltonian one can split those terms that yield the mean energy including 
the mean loss of the system. This justifies the use of passive non-Hermitian 
systems that are inspired by open systems with gain and loss. A gain-loss 
distribution is thus quantum-mechanically equivalent to a system with only loss 
that has the same distribution but with an overall mean loss.

The discussed methods are useful tools to describe passive non-Hermitian 
waveguide systems, allowing for the calculation of analytical and approximate 
solutions as well as providing helpful information on their design and 
implementation.

\acknowledgments
This work was supported by the Deutsche Forschungsgemeinschaft (DFG) through 
grant SCHE 612/6-1.

\appendix

\section{Liouville space ladder operators}
\label{appdx:ladder}

The eigenvectors of $\mathcal{L}$ occurring in 
Sec.~\ref{sec:eigendecomposition} have to be handled with some care because 
$\mathcal{L}$ is non-Hermitian and thus its right and left eigenvectors differ.
This becomes particularly important when calculating the ground states 
$|0\rangle\rangle$ and $\langle\langle 0 |$.
%TODO: check if ground states need to be in boldface!

Let us start with the right ground state and its defining equations given by
Eq.~\eqref{eq:rightGround}. Because the right action of the superoperators 
$P^\pm_i$, $Q^\pm_i$ are known by the definitions of the underlying 
superoperators $L^\pm_i$, $R^\pm_i$, the right ground state and all states 
$|\bm{\alpha}, \bm{\beta} \rangle \rangle$ constructed from it can 
straightforwardly be calculated once the ground state itself is found.
For this we insert a generic operator, e.g.
\begin{equation}
 |0 \rangle \rangle \equiv \hat{\varrho} = \sum_{\bm{n}, \bm{m}} = 
\varrho_{\bm{n}, \bm{m}} | \bm{n} \rangle \langle \bm{m} |
\end{equation}
which results in a set of equations for the coefficients $\varrho_{\bm{n}, 
\bm{m}}$ once the conditions in Eq.~\eqref{eq:rightGround} are calculated.
Solving this set results in the unique solution $|0 \rangle \rangle \equiv | 
\bm{0} \rangle \langle \bm{0} |$.

In order to calculate the left ground state $\langle\langle \bm{0}|$, we first 
have to define the left action of the superoperators $L^\pm_i$, $R^\pm_i$.
For that we use the definition of the adjoint superoperator which is based on 
the inner product $\langle \langle A | B \rangle \rangle = \mathrm{Tr} \left[ 
\hat{A}^\dagger \hat{B} \right]$ endowed on the Liouville space. Using this, we 
obtain for example
\begin{equation}
\langle \langle A | L^+ | B \rangle \rangle = \mathrm{Tr} \left[ 
\hat{A}^\dagger \hat{a}^{\dagger} \hat{B} \right] = \mathrm{Tr} \left[ (\hat{a} 
\hat{A})^\dagger \hat{B} \right] = \langle \langle a A | B \rangle \rangle,
\end{equation}
leading to the left action
\begin{equation}
 \langle \langle A | L^+ = \langle \langle a A |
\end{equation}
and similarly for the other superoperators. The general form of the 
superoperators is 
\begin{gather}
 P^+_i = \sum_k c_{i,k} (L^+_k - R^-_k), \\
 Q^+_i = \sum_k c^*_{i,k} (R^+_k - L^-_k),
\end{gather}
so that we can focus on one of the elements $L^+_k - R^-_k$ or 
$R^+_k -L^-_k$. The defining equations \eqref{eq:leftGround} then become
\begin{equation}
 \langle\langle \bm{0}| (L^+_k - R^-_k) = \langle \langle a A | - \langle 
\langle A a | = 0 ,
\end{equation}
and analagously for the operators $R^+_k - L^-_k$. Clearly, these equations 
are only fulfilled by $A \propto 1_{\mathcal{H}}$. The constant can then be 
determined by choosing the normalisation $\langle \langle \bm{\alpha}, 
\bm{\beta} | \bm{\alpha}, \bm{\beta} \rangle \rangle = 1$.

\section{Structure analysis in the Wei-Norman expansion}
\label{appdx:AlgebraStruc}
The structure analysis used to decompose the algebra spanned by all quadratic 
operators in Sec.~\ref{sec:WeiNorman} can be performed in a systematic manner 
\cite{Gilmore}. The basic tool for this approach is the Cartan-Killing form
\begin{equation}
 \left( A, B \right)_\text{CK} = \mathrm{Tr} \left( \mathcal{R} (A) \mathcal{R} (B) \right)
\end{equation}
with the regular representation \eqref{eq:defRegRepr}, and where
$A$, $B$  are elements of a Lie algebra $\mathfrak{g}$. Applied to an element 
$Z \in \mathfrak{g}$, the Cartan-Killing form $\left( Z, Z \right)_\text{CK}$ 
can either be positive-definite, negative-definite or indefinite. All elements 
that result in an indefinite Cartan-Killing form, i.e. 
$\left(Z,Z\right)_\text{CK} = 0$, together form the maximally solvable 
subalgebra. The rest is split into a compact algebra for which 
$\left(Z,Z\right)_\text{CK}<0$ and a noncompact algebra with 
$\left(Z,Z\right)_\text{CK} > 0$.
This separation into compact and noncompact algebras becomes physically 
relevant because a closed algebra results in a real spectrum.
In the case of the $\mathcal{PT}$-symmetric coupler as discussed
in Sec.~\ref{sec:eigendecomposition}, this would mean that the Cartan-Killing 
form is negative definite in the unbroken $\mathcal{PT}$-phase and positive in 
the broken $\mathcal{PT}$-phase resulting in a real spectrum apart from the 
overall loss factor.

With the above scheme, the decomposition in Sec.~\ref{sec:WeiNorman} can be 
computed by hand for a number of modes $N$ that allow for an efficient 
calculation of the required Cartan-Killing form. Another approach is to
examine the commutation relations of the Lie algebra spanned 
by the quadratic operators $\{ L_i^+ L_j^-, R_i^+ R_j^-, L_i^- R_j^- \}$.
Clearly, the operators $L_i^- R_j^-$ that only remove excitations constitute a 
nilpotent subalgebra because
\begin{gather}
 [L_i^- R_j^-,  L_i^+ L_k^- ] = L_i^- R_j^-, \\
 [L_i^- R_j^-,  R_k^+ R_j^- ] = L_i^- R_j^-, \\
 [L_i^- R_j^-, L_k^- R_l^-]   = 0.
\end{gather}
The remaining elements $\{ L_i^+ L_j^-, R_i^+ R_j^- \}$ can be split into two 
subalgebras of only left or right actions. Because both subalgebras are 
isomorphic, we can define one matrix representation for both which in this case 
are matrices $M_{i,j}$ whose only nonvanishing element (equal to $1$) is 
$(i,j)$. This algebra is the general linear algebra 
$\mathfrak{gl}(N,\mathbb{C})$. 
From this we can extract the special linear algebra 
$\mathfrak{sl}(N,\mathbb{C})$, i.e. the algebra of $N\times N$ matrices with 
vanishing trace, by defining new elements $T_i$ on the main 
diagonal, e.g.
\begin{equation}
 T_i = M_{i,i} - M_{i+1,i+1},
\end{equation}
so that $\mathrm{Tr} \,T_i = 0$. This basis change results in one remaining 
element, i.e. $\sum_i M_{i,i}$, that commutes with all other elements 
$M_{i\neq j}$ and $T_i$. As a result, we split the subalgebra into the algebra 
$\mathfrak{sl}(N,\mathbb{C})$ and the commuting Abelian part $\sum_i M_{i,i}$.
This gives directly the decomposition as in Sec.~\ref{sec:WeiNorman}.

\end{document}